# Chapter 1

# Internet of Things: An Overview

*Farzad Khodadadi, Amir Vahid Dastjerdi, and Rajkumar Buyya*

## Abstract

As technology proceeds and the number of smart devices continues to grow substantially, need for ubiquitous context-aware platforms that support interconnected, heterogeneous, and distributed network of devices has given rise to what is referred today as Internet-of-Things. However, paving the path for achieving aforementioned objectives and making the IoT paradigm more tangible requires integration and convergence of different knowledge and research domains, covering aspects from identification and communication to resource discovery and service integration. Through this chapter, we aim to highlight researches in topics including proposed architectures, security and privacy, network communication means and protocols, and eventually conclude by providing future directions and open challenges facing the IoT development.

## 1.1 Introduction

After four decades from the advent of Internet by ARPANET[1], the term "Internet" refers to vast category of applications and protocols built on top of sophisticated and interconnected computer networks, serving billions of users around the world in 24/7 fashion. Yet, we are at the beginning of an emerging era where ubiquitous communication and connectivity is not a dream or challenge any more. Subsequently, the focus has shifted towards seamless integration of people and devices to converge physical realm with human-made virtual environments, creating the so called Internet-of-Things (IoT) utopia.

A closer look at this phenomenon reveals two important pillars of IoT; "Internet" and "Things" that require more clarification. While it seems that every object capable of connecting to Internet will fall into the "Things" category, this notation is used to encompass more generic set of entities, including smart devices, sensors, human beings, and any other object that is aware of its context and is able to communicate with other entities, making it accessible at any time anywhere. This implies that objects are required to be accessible without any time or place restrictions.

Ubiquitous connectivity is a crucial requirement of IoT and to fulfil it, applications need to support diverse set of devices and communication protocols, from tiny sensors capable of sensing and reporting a desired factor to powerful back-end servers utilized for data analysis and knowledge extraction. This also requires integration of mobile devices, edge devices like routers and smart hubs, and humans in the loop as controllers.

Initially, Radio-Frequency Identification (RFID) used to be the dominant technology behind IoT development, but with further technological achievements, wireless sensor networks (WSN) and Bluetooth-enabled devices augmented the mainstream adoption of IoT trend. These technologies and IoT applications have been extensively surveyed before[2],[3],[4],[5], however less attention has been given to unique characteristics and requirements of IoT such as scalability, heterogeneity support, total integration, and real-time query processing. To make these required advances bold, this chapter lists IoT challenges and promising approaches by considering recent researches and advances that made in the IoT ecosystem as shown in Figure 1.1. In addition, it discusses emerging solutions based on cloud, fog, and mobile computing facilities. Furthermore, the applicability and integration of cutting-edge approaches like Software Defined Networking (SDN) and containers for embedded and constrained devices with IoT is investigated.

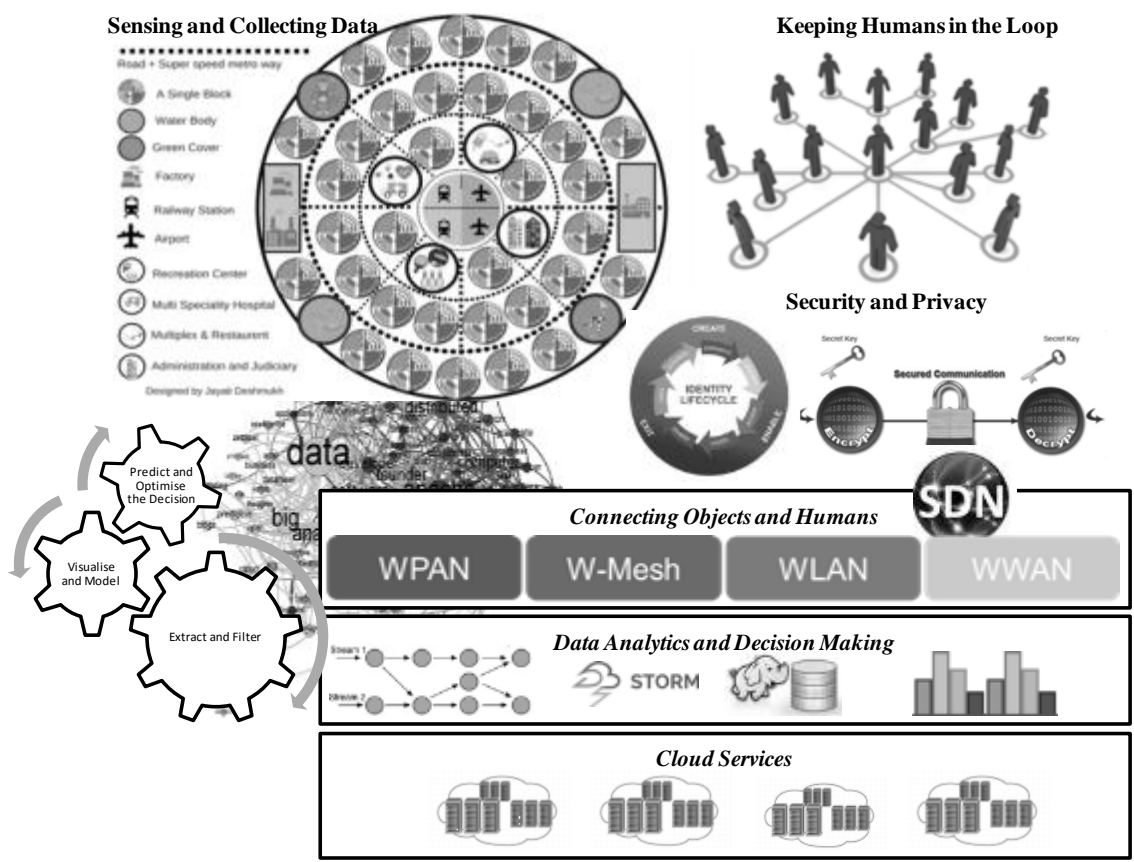

Figure 1.1 : IoT Ecosystem.

# 1.2 Internet-of-Things Definition Evolution

**IoT emergence:** Kevin Ashton is accredited for using the term "Internet-of-Things" for the first time during a presentation in 1999 regarding supply chain management[6]. He believes the "things" aspect of the way we interact and live within the physical world that surround us needs serious reconsideration, due to advances in computing, Internet, and data generation rate by smart devices. At the time, he was an executive director at the MIT's Auto-ID center where contributed to extension of RFID applications into broader domains, which built the foundation for current IoT vision.

**Internet of Everything (IoE):** Since then, many definitions for IoT have been presented, including the definition provided by Gubbi et al.[7] that focuses mostly on connectivity and sensory requirements for entities involved in typical IoT environments. While those definitions reflect IoT's basic requirements, new IoT definitions give more value to the need for ubiquitous and autonomous networks of objects where identification and service integration have an important and inevitable role. For example, Internet of Everything (IoE) is used by Cisco to refer to people, things, and places that can expose their services to other entities[8].

**Industrial IoT (IIoT):** Also referred to as Industrial Internet [87], is another form of IoT applications favoured by big high-tech companies. The fact that machines can perform specific tasks such as data acquisition and communication more accurately than humans has boosted IIoT's adoption. Machine-to-machine (M2M) communication, Big Data analysis, and machine learning techniques are major building blocks when comes to definition of IIoT. This data enables companies to detect and resolve problems faster, thus resulting in overall money and time savings. For instance, in a manufacturing company, IIoT can be used to efficiently track and manage the supply chain, perform quality control and assurance, and lower the total energy consumption.

**Smartness in IoT:** Another characteristic of IoT, which is highlighted in recent definitions, is "smartness". This distinguishes IoT from similar concepts like sensor networks and it can be further categorized into "object smartness" and "network smartness". A smart network is a communication infrastructure characterized by the following functionalities:

- standardization and openness of the communication standards used, from layers interfacing with the physical world (i.e. tags and sensors) up to the communication layers between nodes and with the Internet;
- object addressability (direct IP address) and multi-functionality, i.e. the possibility that a network built for one application (e.g. road traffic monitoring) be available for other purposes (e.g. environmental pollution monitoring or traffic safety) [9].

**Market share:** In addition, definitions draw special attention to potential market of IoT with fast growing rate by having a market value of $44.0 billion in 2011[10]. According to a comprehensive market research conducted by RnRMarketResearch[11] that includes current market size and future predictions, IoT and Machine-to-Machine (M2M) market will be approximately worth $498.92 billion by 2019. Quoting from the same research, the value of IoT market is expected to hit $1423.09 billion by 2020, while Internet of Nano Things (IoNT) playing a key role in future market and holding a value of approximately $9.69 billion by 2020.

Besides all these fantastic and optimistic opportunities, for current IoT to reach the foreseen market, various innovations and progress in different areas are required. Furthermore, cooperation and information sharing between leading companies in IoT such as Microsoft, IBM, Google, Samsung, Cisco, Intel, ARM, Fujitsu, Ecobee Inc, and other smaller businesses and start-ups will boost IoT adoption and market growth.

IoT growth rate with estimated number of active devices until 2018 is depicted in Figure 1.2 [88]. The increase of investment in IoT by developed and developing countries hints at the gradual change in strategy of governments by recognizing IoT's impacts and trying to keep themselves updated as IoT gains momentum. For example, the IoT European Research Cluster (IERC)[1] has conducted and supported several projects about fundamental IoT researches by considering special requirements from end-users and applications. As an example, the project named Internet of Things Architecture (IoT-A) [2] aimed at developing a reference architecture for specific type of applications in IoT and is discussed in more details in Section 1.3. UK government has also initiated a 5 million project on innovations and recent technological advances on IoT[12]. Similarly, IBM in USA[13] have plans to spend billions of dollars on IoT research and its industrial applications. Singapore has also announced its intention to be the first smart nation by investing on smart transport systems, developing the e-government structure, and using surveillance cameras and other sensory devices to obtain data and extract information from them[14].

---

[1] http://www.rfid- in- action.eu/cerp/
[2] http://www.iot-a.eu

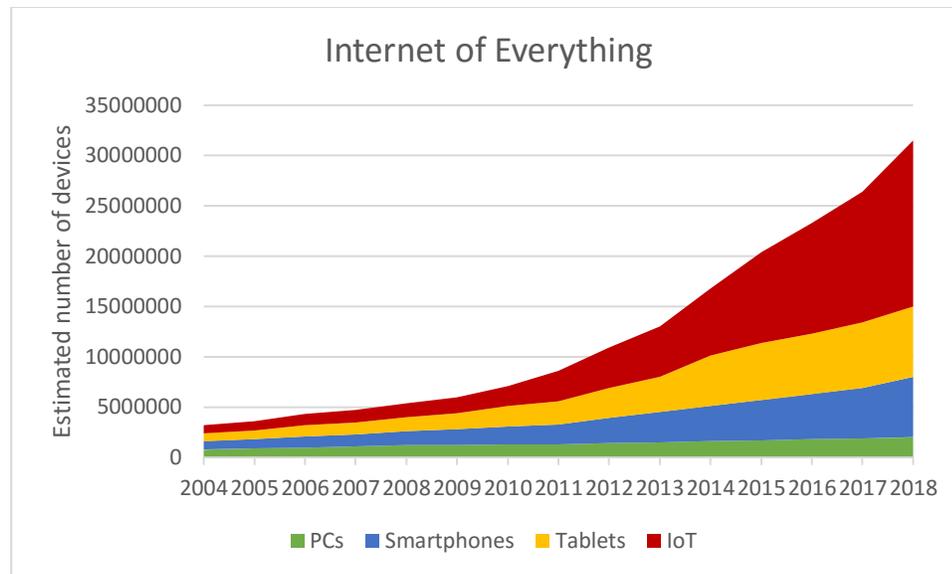

Figure 1.2 : IoT trend forecast [88]

**Human in the loop:** IoT is also identified as an enabler for machine-to-machine, human-to-machine, and human-with-environment interactions. With the increase in number of smart devices and adoption of new protocols such as IPv6, the trend of IoT is expected to shift towards fusion of smart and autonomous network of Internet-capable objects equipped with the ubiquitous computing paradigm. Involving human in the loop [89] of IoT offers numerous advantages to a wide range of applications including emergency management, healthcare, etc. Therefore, another essential role of IoT is building a collaborative system that is capable of effectively responding to an event captured via sensors, by effective discovery of crowds and also successful communication of information across discovered crowds of different domains.

**Improving the quality of life:** IoT is also recognized by the impacts on quality of life and businesses[8] which can revolutionize the way our medical systems and businesses operate via: 1) expanding the communication channel between objects by providing more integrated communication environment where different sensors data such as location, heartbeat , etc can be measured and shared more easily. 2) Facilitating the automation and control process, where administrators can manage each object's status via remote consoles. 3) savings in the overall cost of implementation, deployment, and maintenance, by providing detailed measurements and the ability to check the status of devices remotely.

According to Google Trends, the word "IoT" is used more often than "Internet of Things" since 2004 and after that "web of things" and "Internet of Everything" are the most frequently used words. Quoting the same reference, Singapore and India are the countries with the most regional interest about Internet of Things. This is aligned with the fact that India is estimated to be the world's largest consumer of IoT devices by 2020[15].

# 1.3 IoT Architectures

The building blocks of IoT are sensory devices, remote service invocation, communication networks, and context-aware processing of events and these have been around for many years. However, what IoT tries to picture is a unified network of smart objects and human beings responsible for operating them (if needed) that are capable of universally and ubiquitously communicate with each other.

When talking about a distributed environment, interconnectivity among entities is a critical requirement and IoT is a good example. A holistic system architecture for IoT needs to guarantee flawless operation of its components (reliability is considered as the most import design factor in IoT) and link the physical and virtual realms together. To achieve so, careful consideration is needed in designing failure recovery and scalability.

Additionally, since mobility and dynamic change of location has become an integral part of IoT systems with the widespread use of smart phones, state-of-the-art architectures need to have certain level of adaptability to properly handle dynamic interactions within the whole ecosystem.

Reference architectures and models give a bird-eye view of the whole underlying system, hence their advantage over other architectures relies in providing better and greater level of abstraction which consequently hides specific constraints and implementation details.

Several research groups have proposed reference architectures for IoT[16],[17]. The IoT-A[16] focuses on the development and validation of an integrated IoT network architecture and supporting building blocks, with the objective to be "the European Lighthouse Integrated Project addressing the Internet-of-Things Architecture". IoT-i project, related to the former mentioned IoT-A project, focuses on the promotion of IoT solutions, catching requirements and interests. IoT-i aims to achieve strategic objectives such as: creating a joint strategic and technical vision for the IoT in Europe that encompasses the currently fragmented sectors of the IoT domain holistically, and contributes to the creation of an economically sustainable and socially acceptable environment in Europe for IoT technologies and respective R&D activities.

Figure 1.3 depicts outline of our extended version of a reference architecture for IoT[17]. Different service and presentation layers are shown in this architecture. Service layers include event processing and analytics, resource management and service discovery as well as message aggregation and Enterprise Service Bus (ESB) services built on top of communication and physical layers. API management which is essential for defining and sharing system services and web-based dashboards (or equivalent smartphone applications) for managing and accessing these APIs are also included in the architecture. Due to importance of device management, security and privacy enforcement in different layers, and the ability to uniquely identify objects and control their access level, these components are prestressed independently in this architecture. These components and the related research projects are described in more details throughout this chapter.

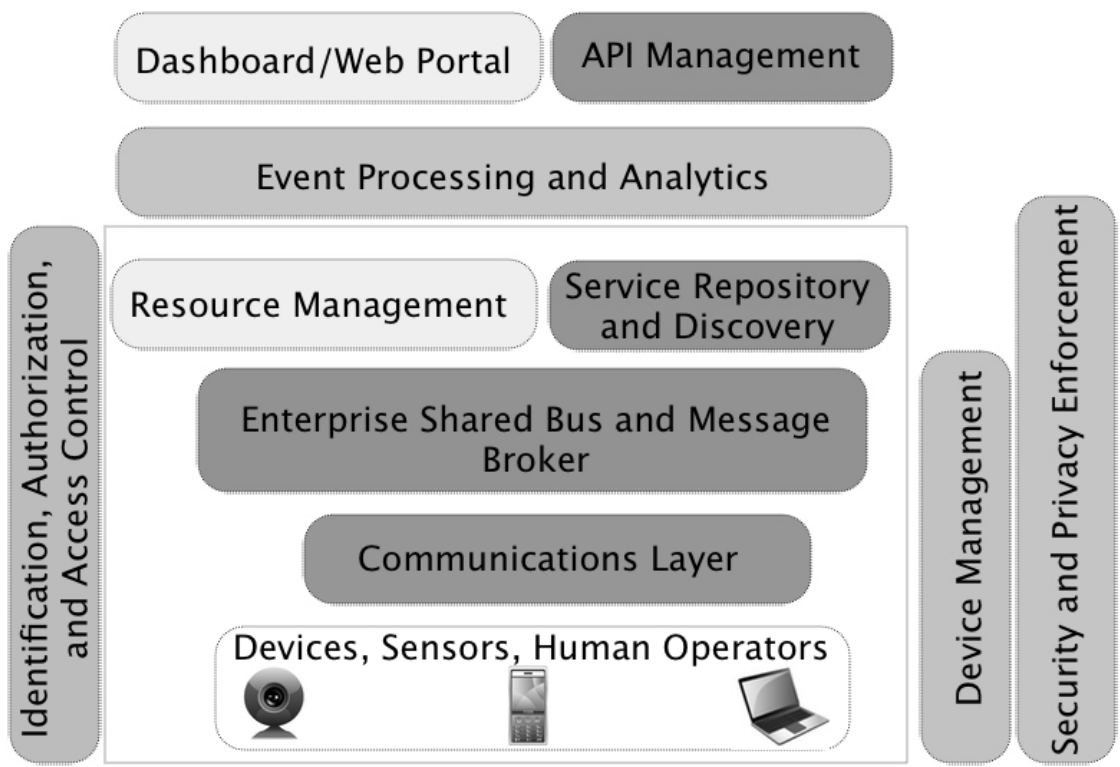

Figure 1.3: A reference architecture for IoT

# 1.3.1 SOA-based Architecture

In IoT, service-oriented architecture (SOA) might be imperative for the service providers and users[18],[19]. SOA ensures the interoperability among the heterogeneous devices [20],[21]. To clarify this, let us consider a generic SOA consists of four layers with distinguished functionalities as below:

- Sensing layer is integrated with available hardware objects to sense the statuses of things;

- Network layer is the infrastructure to support over wireless or wired connections among things;

- Service layer is to create and manage services required by users or applications;

- Interfaces layer consists of the interaction methods with users or applications.

Generally, in such architecture a complex system is divided into subsystems that are loosely coupled and can be reused later (modular decomposability feature), hence providing an easy way to maintain the whole system by taking care of its individual components[22]. This can ensure that in the case of a component failure the rest of the system (components) can still operate normally. This is of immense value for effective design of an IoT application architecture where reliability is the most significant parameter.

SOA has been intensively used in wireless sensor networks, due to its appropriate level of abstraction and advantages pertaining to its modular design[23],[24]. Bringing these benefits to IoT, SOA has the potential to augment the level of interoperability and scalability among the objects in IoT. Moreover, from the user's perspective, all services are abstracted into common sets, removing extra complexity for the user to deal with different layers and protocols[25]. Additionally, the ability to build diverse and complex services by composing different functions of the system (i.e modular composability) though service composition suits the heterogeneous nature of IoT, where accomplishing each task requires a series of service call on all different entities spread across multiple locations [26].

## 1.3.2 API-oriented Architecture

Conventional approaches for developing service-oriented solutions use SOAP and Remote Method Invocation (RMI) as means for describing, discovering, and calling services, however, due to overhead and complexity imposed by these techniques, Web APIs and Representational State Transfer (REST)-based methods introduced as promising alternative solutions. The required resources range from network bandwidth to computational and storage capacity and are triggered by request-response data conversions happening regularly during service calls. Lightweight data exchange formats like JSON can reduce the aforementioned overhead, especially for smart devices and sensors with limited amount of resources, by replacing large XML files used to describe services. This helps in using the communication channel and processing power of devices more efficiently.

Likewise, building APIs for IoT applications helps service provider attract more customers while focusing on the functionality of their products rather than on presentation. In addition, it is easier to enables multi-tenancy by the security features of modern Web APIs such as OAuth, APIs which indeed is capable of boosting an organization's service exposition and commercialization. It also provides more efficient service monitoring and pricing tools than previous service-oriented approaches[27].

To this end, in our previous research we have proposed Simurgh [28] which describes devices, sensors, humans, and their available services using web API notation and API definition languages. Furthermore, a two-phase discovery approach was proposed in the framwork to find sensors that provide desirable services and match certain features, like being in a specific location. Similarly, Elmangoush, et al.[29] proposed a service broker layer (named FOKUS) that exposes set of APIs for enabling shared access to the OpenMTC core. Novel approaches for defining and sharing services in distributed and multi-agent environments like IoT can reduce the sophistication of service discovery in the application development cycle and diminish service call overhead in runtime.

Shifting from Service delivery platforms (SDPs) towards web-based platforms and the benefits of doing so are discussed by Manzalini et al. [30]. Developers and business managers are advised to focus on developing and sharing APIs from the early stage of their application development life cycle, so that eventually by properly exposing data to other developers and end users, an open data environment is created that facilitates collaborative information gathering, sharing, and updating.

# 1.4 Resource Management

Picturing IoT as a big graph with numerous nodes with different resource capacity, selecting and provisioning the resources greatly impacts Quality of Service (QoS) of the IoT applications. Resource management is very important in distributed systems and have been a subject of research for years. What makes resource management more challenging for IoT relies in the heterogeneous and dynamic nature of resources in IoT. Considering large-scale deployment of sensors for a smart city use-case, it is obvious that an efficient resource management module needs consider robustness, fault-tolerance, and scalability, energy efficiency, QoS, and SLA.

Resource management involves discovering and identifying all available resources, partitioning them to maximize a utility function which can be in terms of cost, energy, performance, etc, and finally scheduling the tasks on available physical resources. Figure 1.4 depicts the taxonomy of resource management activities in IoT.

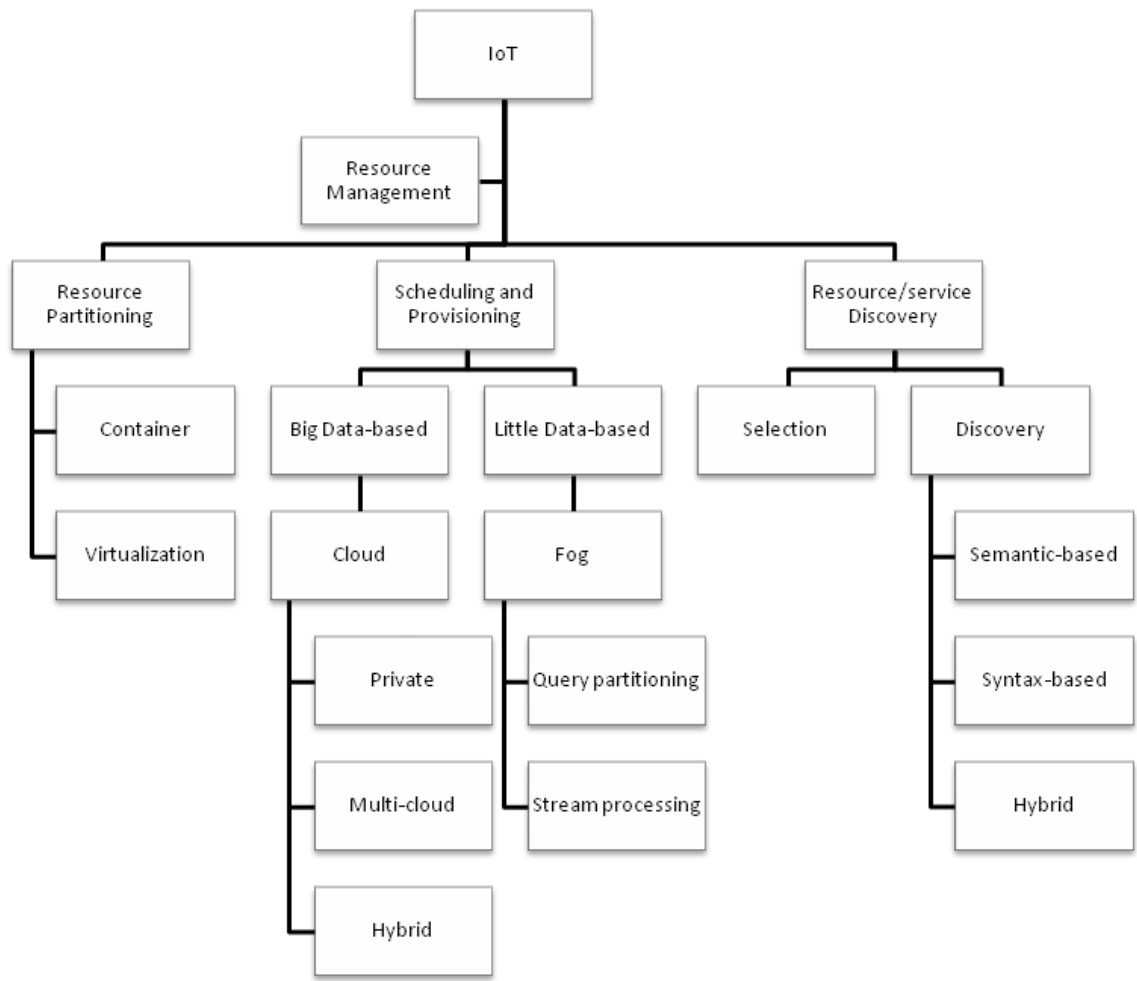

Figure 1.4 : Taxonomy of resource management in IoT

# 1.4.1 Resource Partitioning

The first step for satisfying resource provisioning requirements in IoT is to efficiently partition the resources and gain higher utilization rate. This idea is vastly used in cloud computing via virtualization techniques and commodity infrastructures, however, virtual machines are not the only method for achieving the aforementioned goal. Since the hypervisor, that is responsible for managing interactions between host and guest VMs, require considerable amount of memory and computational capacity, this configuration is not suitable for IoT where devices often have constrained memory and processing power. To address these challenges, the concept of Containers has emerged as a new form of virtualization technology that can match the demand of devices with limited resources. Docker[3] and Rocket [4]are the two most famous container solutions.

Containers are able to provide portable and platform-independent environments for hosting the applications and all their dependencies, configurations, and input/output settings. This significantly reduces the burden of handling different platform-specific requirements when designing and developing applications, hence providing convenient level of transparency for applications architects and developers. In addition, containers are lightweight virtualization solutions that enable infrastructure providers to efficiently utilize their hardware resources by eliminating the need for purchasing expensive hardware and virtualization software packages. Since containers, compared to VMs, require considerably less spin-up time, they are ideal for distributed applications in IoT that need to scale up within a short amount of time.

---

[3] https://www.docker.com/
[4] Available at https://github.com/coreos/rkt

An extensive survey by Gu et al.[91] focuses on virtualization techniques proposed for embedded systems and their efficiency for satisfying real-time application demands. After explaining numerous Xen-based, KVM-based, and microkernel-based solutions that utilise processor architectures such as ARM, authors argue that operating system virtualization techniques, known as container-based virtualization, can bring advantages in terms of performance and security by sandboxing applications on top of a shared OS layer. Linux VServer[92], Linux Containers LXC, and OpenVZ are examples of using OS virtualization in embedded systems domain.

The concept of virtualized operating systems for constrained devices has been further extended to smartphones by providing means to run multiple Android operating systems on a single physical smartphone[93]. With respect to heterogeneity of devices in IoT and the fact that many of them can leverage virtualization to boost their utilisation rate, task-grain scheduling which considers individual tasks within different containers and virtualized environments can potentially challenge current resource management algorithms that view these layers as blackbox[91].

## 1.4.2 Computation Offloading

Code offloading (computation offloading) [90] is another solution for addressing the limitation of available resources in mobile and smart devices. The advantages of using code offloading appear in more efficient power management, less storage requirements, and higher application performance. Several surveys about computation offloading has carefully studied its communication and execution requirements as well as its adaptation

criteria[31],[32],[33], hence here we mention some of the approaches that focus on efficient code segmentation and cloud computing.

Majority of code offloading techniques require the developers to manually annotate the functions required to execute on another device[32]. However, using static code analyzers and dynamic code parsers is an alternative approach that results in better adaptivity in case of network fluctuations and increased latency[34]. Instead of using physical instances, ThinkAir[35] and COMET[36] leverage virtual machines offered by IaaS cloud providers as offloading targets to boost both scalability and elasticity. The proposed combination of VMs and mobile clouds can create a powerful environment for sharing, synchronizing, and executing codes in different platforms.

## 1.4.3 Identification and Resource/Service Discovery

Internet of Things has emerged as a great opportunity for industrial investigations and similarly pursued by research communities, but current architectures proposed for creation of IoT environments lack support for efficient and standard way of service discovery, composition, and their integration in scalable manner[37].

The discovery module in IoT is twofold. First objective is to identify and locate the actual device, which can be achieved by storing and indexing metadata information about each object. The final step is to discover the target service that needs to be invoked.

Lack of effective discovery algorithm can result in execution delays, poor use experience, and runtime failures. As discussed in[38], efficient algorithms that dynamically choose centralized or flooding strategies can help minimize the consumed energy, although, other parameters such as mobility and latency should be factored in to offer a suitable solution for IoT, considering its dynamic nature. In another approach within the fog computing context[39], available resources like network bandwidth and computational and storage capacity metrics are converted to time resources, forming a framework that facilitates resource sharing. Different parameters like energy consumption level, price, and availability of services need to be included in proposing solutions that aim to optimize resource sharing within heterogeneous pool of resources.

The Semantic Web of Things (SWoT) envisions advanced resource management and service discovery for IoT by extending Semantic Web notation and blending it with IoT and Web of Things. To achieve so, resources and their metadata are defined and annotated using standard ontology definition languages such as RDF and OWL. Additionally, search and manipulation of these metadata can be done through query languages like SPARQL. Ruta et al [94]  has adopted the SSN-XG W3C ontology to collect and annotate data from Semantic Sensor Networks (SSN) and by extending the CoAP protocol (discussed in section 1.6) and CoRE Link Format that is used for resource discovery, their proposed solution ranks resources based on partial or full request matching situations.

# 1.5 IoT Data Management and Analytics

While Internet of Things (IoT) is getting momentum as enabling technology for creating a ubiquitous computing environment, special considerations are required to process huge amount of data originating from and circulating in such a distributed and heterogeneous environment. To this extent, Big Data related procedures such as data acquisition, filtering, transmission, and analysis have to be updated to match the requirements of IoT data deluge.

Generally, Big Data is characterized by 3Vs, namely velocity, volume, and variety. Focusing on individual or combination of the three mentioned Big Data dimensions has lead to the introduction of different data processing approaches. Batch Processing and Stream Processing are two major methods used for data analysis. Lambda Architecture [40] is an exemplary framework proposed by Nathan Marz to handle Big data processing by focusing on multi-application support, rather than data processing techniques. It has three main layers that enable the framework to support easy extensibility through extension points, scale-out capabilities, low latency query processing, and the ability to tolerate human and system faults. From a top-down view, first layer is called "Batch Layer" and hosts the master dataset and batch views where pre-computed queries are stored. Next is the "Serving Layer" which adds dynamic query creation and execution to the batch views by indexing and storing them, and finally, the "Speed Layer" captures and processes recent data for delay-sensitive queries.

Collecting and analysing the data circulating in IoT environment is where the real power of IoT resides [41]. To this end, applications utilize pattern detection and data mining techniques to extract knowledge and make smarter decisions. One of the key limitations in using currently developed data mining algorithms lies in the inherent centralized nature of these algorithms which drastically affects their performance and makes them unsuitable for IoT environments that are meant to be geographically distributed and heterogeneous. Distributed anomaly detection techniques that process multiple streams of data concurrently to detect outliers have been well studied in the literature [42]. A comprehensive survey of data mining researches in IoT has been conducted by Tsai et al.[44] and includes details about various classification, clustering, knowledge discovery in databases (KDD), and pattern mining techniques. Nevertheless, new approaches like ellipsoidal neighbourhood factor outlier [43] that can be efficiently implemented on constrained devices are not fully benchmarked in respect to different configurations of their host devices.

## 1.5.1 IoT and the Cloud

Cloud computing due to its on-demand processing and storage capabilities can be used to analyse data generated by IoT objects in batch or stream format. Pay-as-you-go model adopted by all cloud providers has reduced the price of computing, data storage, and data analysis, creating a streamlined process for building IoT applications. With cloud's elasticity, distributed Stream Processing Engines (SPEs) can implement important features such as fault-tolerance and auto-scaling for bursty workloads.

IoT application development in clouds has been investigated in number of researches. Alam et al.[45] proposed a framework that supports sensor data aggregation in cloud-based IoT context. The framework is an SOA-based in event-driven and defines and a benefits from a semantic layer that is responsible for event processing and reasoning. Similarly, Li et al.[46] proposed a Platform as a Service (PaaS) solution for deployment of IoT applications. The solution is multi-tenant and a virtually isolated service is provided for users that can be customized to their IoT devices while sharing the underlying cloud resources with other tenants.

Nastic et al.[47] proposed PatRICIA, a framework that provides a programming model for development of IoT applications in the cloud. PatRICIA proposes new abstraction layer that is based on the concept of Intent-based programming. Parwekar[48] discussed the importance of identity detection devices in IoT and proposed a service layer to demonstrate how a sample tag-based acquisition service can be defined in the cloud. A simple architecture for integrating machine to machine (M2M) platform, network, and data layers has also been proposed. Focusing on the data aspect of IoT, in our previous research we proposed an architecture based on Aneka by adding support for data filtering, multiple simultaneous data source selection, load balancing, and scheduling[49].

IoT applications can harness cloud services and use the available storage and computing resources to meet their scalability and compute-intensive processing demands. Most of current design approaches for integrating cloud with IoT are based on a three tier architecture where the bottom layer consists of IoT devices, middle layer is the cloud provider, and top layer hosts different applications and high-level protocols. However, using this approach to design and integrate cloud computing with an IoT middleware limits the practicality and fully utilization of cloud computing in scenarios where minimizing end-to-end delay is the goal. For example in online game streaming where perceived delay is an important factor for user satisfaction, a light and context-aware IoT middleware [50] that smartly selects nearest Content Distribution Network (CDN) can significantly reduce the overall jitter.

## 1.5.2 Real-time Analytics in IoT and Fog Computing

Current data analytics approaches mainly focus on dealing with Big Data, however, processing data generated from millions of sensors and devices in real time is a more challenging[51]. Proposed solutions that only utilize cloud computing as processing or storage backbone are not scalable and cannot address the latency constraints of real-time applications. Real-time processing requirements and the increase in computational power of edge devices like routers, switches, and access points lead to the emergence of Edge Computing paradigm.

Edge layer contains the devices that are in closer vicinity to the end user than the application servers and can include smartphones, smart TVs, network routers, etc. Processing and storage capability of these devices can be utilized to extend the advantages of using cloud computing by creating another cloud, known as Edge Cloud, near application consumers in order to: decrease networking delays, save processing or storage cost, perform data aggregation, and avoid sensitive data leaving the local network[52].

Similarly, Fog Computing is a term coined by Salvatore Stolfo[53] and applies to an extension of cloud computing that aims to keep the same features of Cloud such as networking, compute, virtualization, and storage, but also meet the requirements of applications that demand low latency, specific QoS requirements, Service Level Agreement (SLA) considerations, or any combination of them[54]. Moreover, these extensions can ease application development for mobile applications, Geo-distributed applications such as wireless sensor networks, and large-scale systems used for monitoring and controlling other systems, such as surveillance camera networks[55],[56].

A comparison of Cloud and Fog features is presented in Table 1.1 and Figure 1.5 shows a general architecture for using cloud and fog computing together.

Table 1.1: Cloud versus Fog

| | Fog | Cloud |
|---|---|---|
| Response time | Low | High |
| Availability | Low | High |
| Security level | Medium to hard | Easy to medium |
| Service focus | Edge devices | Network/enterprise core services |
| Cost for each device | Low | High |
| Dominant architecture | Distributed | Central/distributed |
| Main content generator-consumer | Smart Devices-humans and devices | Humans-end devices |

Stonebraker et al. [57] pointed that the following requirements should be fulfilled in an efficient real-time stream processing engine (SPE):

1) data fluidity, which refers to processing data on-the-fly without need for costly data storage,

2) handling out-of-order, missing, and delayed streams,

3) having repeatable and deterministic outcome after processing series or bag of streams,

4) keeping streaming and stored data integrated by using embedded database systems,

5) assuring high-availability using real-time failover and hot backup mechanisms,

6) supporting auto-scaling and application partitioning.

To harness the full potential of Fog computing for applications demanding real-time processing, researcher can look into necessary approaches and architectures to fulfil the above-mentioned requirements.

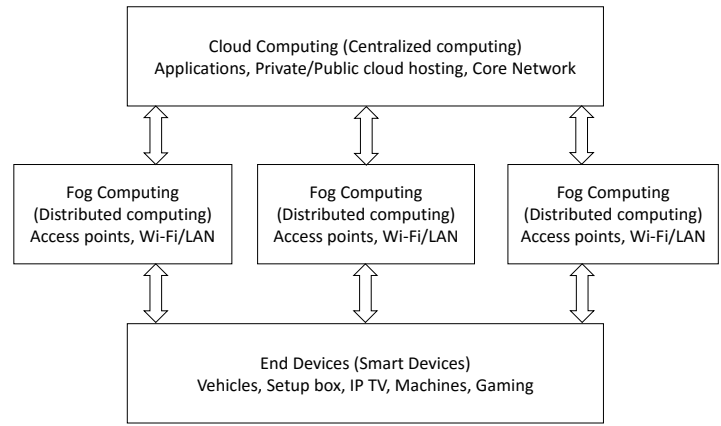

Figure 1.5: Typical Fog computing architecture

# 1.6 Communication Protocols

From the network and communication perspective, IoT can be viewed as aggregation of different networks including mobile networks (3G, 4G, CDMA, etc), WLANs, Wireless Sensor Networks (WSN) and Mobile Adhoc Networks (MANET)[18].

Seamless connectivity is a key requirement for IoT. Network communication speed, reliability, and connection durability will impact overall IoT experience. With the emergence of high-speed mobile networks like 5G and higher availability of local and urban network communication protocols such as Wi-Fi, Bluetooth, and WiMax, creating an interconnected network of objects seems feasible, however dealing with different communication protocols that link these environments is still challenging.

# 1.6.1 Network Layer

Based on the devices specification (memory, CPU, storage, battery life), the communication means and protocols vary. However, the commonly used communication protocols and standards are listed below:

- RFID (e.g. ISO 18000 series that come with 5 classes and 2 Generations and cover both active and passive RFID tags)

- IEEE 802.11 (WLAN), IEEE 802.15.4 (ZigBee), Near Field Communication (NFC), IEEE 802.15.1 (Bluetooth)

- Low power Wireless Personal Area Networks (6LoWPAN) standards by IEFT

- Machine to Machine (M2M) protocols such as MQTT and CoAP

- IP layer technologies such as IPv4, IPv6, etc.

More elaboration on the above mentioned network layer communication protocols is available in [58] and a breakdown of layers in IoT communication stack that these protocols will operate is shown in Figure 1.6.

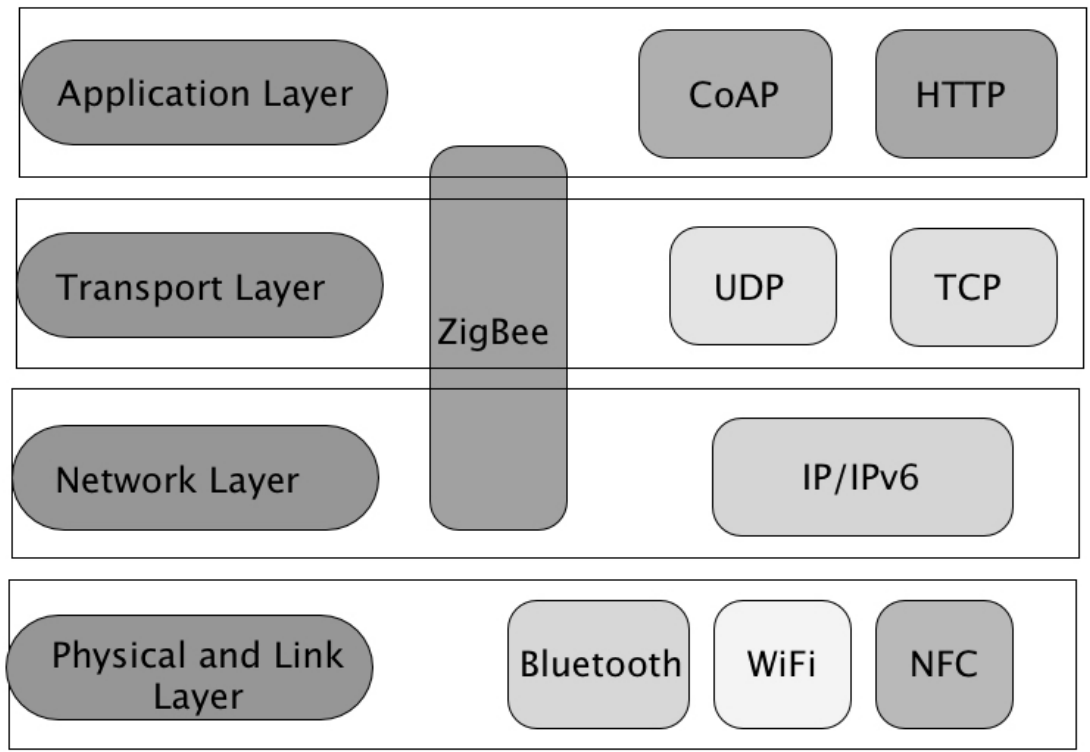

Figure 1.6: Use of various protocols in IoT communication layers

## 1.6.2 Transport and Application Layer

Segmentation and poor coherency level, which are results of pushes from individual companies to maximize their market share and revenue, has made developing IoT applications cumbersome. Universal applications that require one-time coding and can be executed on multiple devices are the most efficient.

Protocols in IoT can be classified into three categories:

1) general-purpose protocols like IP and SNMP that has been around for many years and are vastly used to manage, monitor, configure network devices, and establish communication links ;

2) lightweight protocols such as CoAP that has been developed to meet the requirements of constrained devices with tiny hardware and limited resources;

3) device or vendor specific protocols and APIs that usually require certain build environment and toolset.

Selecting the right protocols at the development phase can be challenging and complex as factors such as future support, ease of implementation, and universal accessibility have to be considered. Additionally, thinking of other aspects that will affect the final deployment and execution, like required level of security and performance, will add to the sophistication of protocol selection stage. Lack of standardization for particular applications and protocols is another factor that increases the risk of poor protocol selection and strategic mistakes that are more expenses to fix in the future. Yet abother challenge is insufficient documentation for some protocols sensors and smart devices limits their usage in IoT.

Table 1.2 summarizes the characteristics of major communication protocols in IoT, while also comparing their deployment topology and environments.

Table 1.2: IoT communication protocols comparison

| Protocol Name | Transport Protocol | Messaging Model | Security | Best Use cases | Architecture |
|---|---|---|---|---|---|
| AMPQ | TCP | Publish/Subscribe | High-Optional | Enterprise Integration | P2P |
| CoAP | UDP | Request/Response | Medium-Optional | Utility field | Tree |
| DDS | UDP | Publish/Subscribe and Request/Response | High-Optional | Military | Bus |
| MQTT | TCP | Publish/Subscribe and Request/Response | Medium-Optional | IoT messaging | Tree |
| UPnP | - | Publish/Subscribe and Request/Response | None | Consumer | P2P |
| XMPP | TCP | Publish/Subscribe and Request/Response | High-Compulsory | Remote management | Client server |
| ZeroMQ | UDP | Publish/Subscribe and Request/Response | High-Optional | CERN | P2P |

Machine to Machine (M2M) communication aims to enable seamless integration of physical and virtual objects into larger and geographically distributed enterprises by eliminating the need for human intervention. To achieve so, enforcing harmony and collaboration among different communication layers (physical, transport, presentation, application) and approaches used by devices for message storage and passing can be challenging [59].

Publish/subscribe model is a common way of exchanging messages in distributed environments and because of simplicity, it has been adopted by popular M2M communication protocols like MQTT. In dynamic scenarios where nodes join or leave the network frequently and handoffs are required to keep the connections alive, publish/subscribe model is efficient. This is because of using push-based notifications and maintaining queues for delayed delivery of messages.

On the other hand, protocols like HTTP/REST and CoAP only support the request/response model in which pulling mechanism is used to fetch new messages from the queue. CoAP also uses IPv6 and 6LoWPAN protocols in its network layer to handle node identification. Ongoing efforts are still being made to merge these protocols and standardize them as to support both publish/subscribe and request/response models[60],[61].

# 1.7 Internet of Things Applications

IoT promises an interconnected network of uniquely identifiable smart objects. This infrastructure creates the necessary backbone for many interesting applications that require seamless connectivity and addressability between their components. The range of IoT application domain is wide and encapsulates applications from home automation to more sophisticated environments such as smart cities and e-government.

Industry-focused applications include logistics and transportation[62], supply chain management[63], fleet management, aviation industry, and enterprise automation systems. Healthcare systems, smart cities and buildings, social IoT, and smart shopping are few examples of applications that try to improve the daily life of individuals, as well as the whole society. Disaster management, environmental monitoring, smart watering and optimizing energy consumption through smart grids and smart metering are examples of applications that focus on environment.

In a broader magnitude, Asin and Gascon[64] classified 54 different IoT applications under the following categories: smart environment, smart cities, smart metering, smart water, security and emergencies, retail, logistics, industrial control, smart agriculture, smart animal farming, domestic and home automation, and eHealth. For further reference, Kim et al.[65] have surveyed and classified researches about IoT applications based on application domain and target user groups.

In this section we present categorization of enterprise IoT applications based on their usage domain. These applications usually fall into the following three categories: 1) Monitoring and actuating, 2) Business process and data analysis, 3) Information gathering and collaborative consumption. The rest of this section is dedicated to characteristics and requirements of each category.

# 1.7.1 Monitoring and actuating

Monitoring devices via APIs can be helpful in multiple domains. The APIs can report power usage, equipment performance, sensors status, and perform actions upon sending pre-defined commands. Real-time applications can utilise these features to report current system status, while managers and developers have the option to freely call these APIs without the need of physically accessing the devices. Smart metering, and in a more distributed form, smart grids can help in identifying production or performance defects via application of anomaly detection on the collected data and thus increase the productivity. Likewise, incorporating IoT in building to or even in the construction process [66] helps in moving towards green solutions, saving energy ,and consequently minimising operation cost.

Another area that has been under focus by researchers is applications targeting smart homes that mainly target energy saving and monitoring. Home monitoring and control frameworks like the ones developed by Verizon[67] and Boss support different communication protocols (Wi-Fi, Bluetooth, etc) to a create an interconnected network of objects that can control desired parameters and change configurations based on user's settings.

# 1.7.2 Business process and data analysis

Riggins et al. [68] categorized level of IoT adoption through Big Data analytics usage to the following categories:

1) *Society level* where IoT mainly influences and improves government services by reducing cost and increasing government transparency and accountability,

2) *Industry level* in which manufacturing, emergency services, retailing, and education have been studied as examples,

3) *Organizational level* in which IoT can bring same type of benefits as those mentioned in society level,

4) *Individual level* where daily life improvements and individual efficiency and productivity growth are marked as IoT benefits.

The ability to capture and store vast amounts of individual data has brought opportunities to healthcare applications. Patients' data can be captured more frequently, using wearable technologies such as smart watches, and can be published over internet. Later, data mining and machine learning algorithms are used to extract knowledge and patterns from the raw data and archive these records for future references. Healthsense eNeighbor developed by Humana is an example of a remote controlling system that uses sensors deployed in houses to measure frequent daily activities and heath parameters of occupants. The collected data is then analysed to forecast plausible risks and produce alerts prevent incidents [69]. Privacy and security challenges are two main barriers that refrain people and industries from embracing IoT in the healthcare domain.

# 1.7.3 Information gathering and collaborative consumption

Social Internet of Things (SIoT) is where IoT meets social networks and to be more precise, it promises to link objects around us with our social media and daily interaction with other people, making them look smarter and more intractable. SIoT concept, motivated by famous social media like Facebook and Twitter, has the potential to affect many people's life style. For example, social network is helpful for evaluation of trust of crowds involved in an IoT processes. Another advantages is using the humans and their relationships, communities, and interactions for effective discovery of IoT services and objects [95].

Table 1.3 contains a list of past and present open source projects regarding IoT development and its applications

Table 1.3: List of IoT-related projects

| Name of project/product | Area of focus |
| --- | --- |
| Tiny OS | Operating System |
| Contiki | Operating System |
| Mantis | Operating System |
| Nano-RK | Operating System |
| LiteOS | Operating System |
| FreeRTOS | Operating System |
| RIOT | Operating System |
| Wit.AI | Natural Language |
| Node-RED | Visual Programming Toolkit |
| NetLab | Visual Programming Toolkit |
| SensorML | Modeling and Encoding |
| Extended Environments Markup Language (EEML) | Modeling and Encoding |
| ProSyst | Middleware |
| MundoCore | Middleware |

| Gaia | Middleware |
|------|-----------|
| Ubiware | Middleware |
| SensorWare | Middleware |
| SensorBus | Middleware |
| OpenIoT | Middleware and development platform |
| Koneki | M2M Development Toolkit |
| MIHINI | M2M Development Toolkit |

# 1.8 Security

As adoption of IoT continues to grow, attackers and malicious users are shifting their target from servers to end devices. There are several reasons for this, first in terms of physical accessibility, smart devices and sensors are far less protected than servers and having physical access to a device gives the attackers privilege to penetrate with less hassle. Second, the number of devices that can be compromised are way more than number of servers. Moreover, since devices are closer to the users, security leads to leak of valuable information and has catastrophic consequences. Finally, due to heterogeneity and distributed nature of IoT, patching process is more consuming, thus opening the door for attackers[71].

In an IoT environment, resource constraints are the key barrier for implementing standard security mechanisms in embedded devices. Furthermore, wireless communication used by majority of sensor networks is more vulnerable to eavesdropping and man-in-the-middle (proxy) attacks.

Cryptographic algorithms need considerable bandwidth and energy to provide end-to-end protection against attacks on confidentiality and authenticity. Solutions have been proposed in RFID[72],[73] and wireless sensor network[74] context to overcome aforementioned issues by considering light cryptographic techniques. With regards to constrained devices, symmetric cryptography is applied more often as it requires less resources, however public key cryptography in the RFID context has been also investigated[75].

Wireless sensor networks with RFID tags and their corresponding readers were the first infrastructure for building IoT environments and even now, many IoT applications in logistics, fleet management, controlled farming, and smart cities rely on these technologies. Nevertheless, these systems are not secure enough and are vulnerable to various attacks from different layers. A survey by Borgohain et al. [76] investigate these attacks, but less attention is given to solutions and counter-attack practices.

## 1.9 Identity Management and Authentication

When talking about billions of connected devices, methods for identifying objects and setting their access level play an important role in the whole ecosystem. Consumers, data sources, and service providers are essential parts of IoT, identity management and authentication methods applied to securely connect these entities affects both the amount of time required to establish trust and the confidence degree[4]. IoT's inherent features like dynamism and heterogeneity require specific consideration when defining security mechanisms. For instance, in Vehicular Networks (VANETs), cars regularly enter and leave the network due to their movement speed, thus not only cars need to interact and exchange data with access points and sensors along the road, but also they need to communicate with each other and form a collaborative network.

Devices or objects in IoT have to be uniquely identified. There are various mechanisms such as ucode which generates 128 bit codes and can be used in active and passive RFID tags and Electric Product Code (EPC) which creates unique identifiers using Uniform Resource Identifier (URI) codes[77],[78]. Being able to globally and uniquely identify and locate objects decreases the complexity of expanding the local environment and linking it with the global markets[76].

It is common for IoT sensors and smart devices to share the same geographical coordinates and even fall into same type or group, hence identity management can be delegated to local identity management systems. In such environments, local identity management systems can enforce and monitor access control policies and establish trust negotiations with external partners. Liang et al. [79] investigated security requirements for multimedia applications in IoT and proposed an architecture that supports traffic analysis and scheduling, key management, watermarking, and authentication. Context-aware pairing of devices and automatic authentication is another important requirement for dynamic environments like IoT. Solutions that implement zero-interaction approach[80] to create simpler yet more secure procedure for creating ubiquitous network of connected devices can considerably impact IoT and its adoption.

# 1.10 Privacy

According to the report published by IDC and EMC on December 2012[81], the size of digital universe containing all created, replicated, and consumed digit data will be roughly doubled each two year, hence, forecasting its size to be 40,000 Exabytes till 2020, compared to 2.837 Exabytes for 2012. Additionally, sourced from

statisticbrain.com, the average cost of storage for hard disks has dropped from $437,500 per Gigabyte in 1980 to $0.05 per Gigabyte in 2013. These statistics show the importance of data and the fact that it is easy and cheap to keep user's data for a long time and follow the guideline of harvesting as much data as possible and using it when required.

Data generation rate has drastically increased in recent years and consequently concerns about secure data storage and access mechanisms has  be taken more seriously. With sensors capable of sensing different parameters such as users' location, heartbeat, and motion, data privacy will remain a hot topic to ensure users have control over the data they share and the people who have access to these data.

In distributed environments like IoT, preserving privacy can be achieved by either following a centralized approach or by having each entity manage its own inbound/outbound data, a technique known as privacy-by-design[76]. Considering the latter approach, since each entity can access only chunks of data, distributed privacy preserving algorithms have been developed to handle data scattering and their corresponding privacy tags[82]. Privacy enhancing technologies[83], [84] are good candidates for protecting collaborative protocols. In addition, to protect sensitive data, rapid deployable enterprise solutions that leverage containers on top of virtual machines can be used[85].

## 1.11 Standardization and Regulatory Limitations

Standardization and the limitation caused by regulatory policies have challenged the growth and adoption rate of IoT and can be potential barriers in embracing the technology. Defining and broadcasting standards will ease the burden of joining IoT environments for new users and providers. Additionally, interoperability among different components, service providers, and even end users will be greatly influenced in a positive way, if pervasive standards are introduced and employed in IoT[86].

Even though more organizations and industries make themselves ready to embrace and incorporate IoT, increase in IoT growth rate will cause difficulties for standardization. Strict regulations about accessing radio frequency levels, creating sufficient level of interoperability among different devices, authentication, identification, authorization, and communication protocols are all open challenges facing IoT standardization. Table 1.4 contains a list of organizations that has worked towards standardizing technologies used within IoT context or those specifically created for IoT.

Table 1.4: IoT standards

| Organization Name | Outcome |
|---|---|
| Internet of Things Global Standards Initiative (IoT-GSI) | JCA-IoT |
| Open Source Internet of Things (OSIOT) | Open Horizontal Platform |
| IEEE | 802.15.4 standards, developing a reference architecture |
| Internet Engineering Task Force (IETF) | Constrained RESTful Environments (CoRE), 6LOWPAN, Routing Over Low power and Lossy networks (ROLL), IPv6 |
| The World Wide Web Consortium (W3C) | Semantic Sensor Net Ontology, Web Socket , Web of Things |
| XMPP Standards Foundation | XMPP |
| Eclipse Foundation | Paho project, Ponte project, Kura, Mihini/M3DA, Concierge |
| Organization for the Advancement of Structured Information Standards | MQTT, AMPQ |

# 1.12 Conclusions

Internet-of-Things has emerged as a new paradigm aiming at providing solutions for integration, communication, data consumption and analysis of smart devices. To this end, connectivity, interoperability, and integration are inevitable parts of IoT communication systems. While IoT, due to its highly distributed and heterogeneous nature, is comprised of many different components and aspects, providing solutions to integrate this environment and hide its complexity from the user side is inevitable. Novel approaches that utilize SOA architecture and API definition languages to service exposition , discovery, and composition will have huge impact in adoption and proliferation of the future IoT vision.

In this paper, different building blocks of IoT such as sensors and smart devices, M2M communication, and the role of humans in future IoT scenarios are elaborated and investigated. Many challenges ranging from communication requirements to middleware development still remain open and need further investigations. We highlighted these shortcomings and provided typical solutions and draw guidelines for future researches in this area.